\documentclass{iaus}
\usepackage{graphicx}

\title[IAU 283 Symp. - Extragalactic PNs] 
      {Planetary Nebulae Populations in External Galaxies}

\author[Magda Arnaboldi]   
{Magda Arnaboldi$^{1,2}$}

\affiliation{$^1$European Southern Observatory, \\ Karl-Schwarzchild-Strasse 2, 85748 Garching bei M\"unchen, Germany\\ email: {\tt marnabol@eso.org} \\
[\affilskip]
  $^2$INAF, Osservatorio Astronomico di Pino Torinese, \\ I-10025 Pino Torinese
, Italy }

\pubyear{2012}
\volume{283}  
\pagerange{1--8}
\jname{Planetary Nebulae - An Eye to the future}
\editors{A. Machado, L. Stanghellini, \&  D. Sch\"onberner eds.}

\begin{document}

\maketitle

\begin{abstract}
This review highlights the properties of the planetary nebulae in
external galaxies as tracers of light, of the stellar population
properties, and of the distances and kinematics of the parent
galaxies. Recent results on the kinematics of the outer regions in
giant elliptical galaxies and on the luminosity specific PN numbers
(the $\alpha$ parameter) in these systems are presented, based on
current surveys of planetary nebulae with the Planetary Nebulae
Spectrograph (PN.S) and other instruments. Finally a brief discussion
is given of planetary nebulae as tracers of the diffuse light in the
nearby clusters, such as Virgo and Hydra~I.

\end{abstract}

\firstsection 
\section{Introduction}\label{intro}

Planetary Nebulae (PNs) are stars that evolve from the asymptotic
giant branch (AGB) to their final destiny as white dwarfs. The
majority of stars between $1$ and $8\, M_\odot$ evolve through a PN
phase. As a consequence, in old stellar populations as those observed
in bulges and early-type galaxies, most stars will reach the post AGB
phase and a fraction of those stars will go through a PN phase, before
ending their lives as white dwarfs.

In a PN, the central star emits most of the light in the UV part of
the spectrum. The nebular shell of the PN is able to convert the
UV ionizing photons into various line emissions from the UV to the
optical and down to the NIR. Up to $15\%$ of the UV emitted energy by
the central stars is re-emitted in the [OIII] $\lambda$ 5007 \AA\ line, the
brightest optical emission of a PN (\cite[Dopita et al. 1992]{Dopita+92}).

We can integrate the whole [OIII] flux emitted from a PN and derive
the $m_{5007}$ magnitude defined as
\begin{equation}
m_{5007} = -2.5 \log F_{5007} - 13.74
\end{equation}
(\cite[Jacoby 1989]{Jac89}).  For a PN population observed in an
external galaxy, we can derive the PN luminosity function (PNLF) for
that population, which is empirically shown to follow an
analytical formula
\begin{equation}
N(m_{5007}) = C\times e^{0.307m_{5007}} \times [ 1- e^{3(m^*-m_{5007})}]
\label{PNLF}
\end{equation}
(\cite[Ciardullo \etal\ 1998]{Ciardullo+98}) where $m^*$ is the
apparent magnitude of the bright cut-off. This analytic formula
combines the observed behavior at the bright end, which is believed to
originate from the most massive surviving stellar cores
(\cite[Ciardullo \etal\ 1989]{Ciardullo+89}), and the slow PN fading
rate caused by the envelope expansion at the faint end (\cite[Heinze
  \& Westerlund 1963]{HenWest1963}).  When this equation is integrated
down 8 mag from $m^*$, it provides the total number $N_{PN}$ of PNs
associated with the total luminosity of a given galaxy. The magnitude
range of 8 mag from $m^*$ allows us to account for the faintest known
Galactic planetary nebula (\cite[Ciardullo
  \etal\ 1989]{Ciardullo+89}).

In external galaxies, PN are most often observed using a narrow band
filter centered on the redshifted wavelength of the brightest emission
line, the PN [OIII] $\lambda$ 5007 \AA\ line.  In galaxies at
distances larger than $2$ Mpc, the PN emissions are spatially
unresolved.

{\underline{\it Planetary Nebulae trace light}}. The PN population in
an external galaxy is expected to trace light because the
luminosity-specific stellar death rate ({\rm B}) should be independent
of the precise state of the underlying stellar population
(\cite[Renzini \& Buzzoni 1986]{RenBuz86}). \cite[Buzzoni
  \etal\ (2006)]{BAC2006} computed the specific evolutionary flux {\rm
  B} for simple stellar populations using different Initial Mass
Functions (IMFs) and different ages, and measured a variation of {\rm
  B} by at most a factor of $2$.  We can write the total number of
PNs, $N_{PN}$, associated with a given bolometric luminosity from a
simple stellar population as
\begin{equation}
N_{PN} = {\rm B}\times L_{TOT}\times \tau_{PN}
\end{equation}
where 
\begin{itemize}
\item $N_{PN}$ is the total number of PNs associated with the parent
  stellar population. $N_{PN}$ is obtained by integration of the PNLF
  $8$ mags down the bright cut off;
\item {\rm B} is the luminosity-specific stellar death rate or
  specific evolutionary flux;
\item $L_{TOT}$ is the total bolometric luminosity of the parent
  simple stellar population;
\item $\tau_{PN}$ is the PN visibility lifetime, i.e. the time for the
  nebula to be detectable in either [OIII] or H$\alpha$ surveys. The
  upper limit of $\tau_{PN}$ is $30000$ yr, that is the time required
  for a PN nebular envelope to fade entirely because of its expansion
  (\cite[Heinze \& Westerlund 1963]{HenWest1963}).
\end{itemize}
We can define the luminosity specific PN number, $\alpha$-parameter
for short, as
\begin{equation} 
\alpha = \frac{N_{PN}}{L_{TOT}}= {\rm B}\times \tau_{PN}
\label{taupn}
\end{equation} 
which provides us with a population averaged-measurement of the PN
visibility lifetime $\tau_{PN}$ for a PN population.  Because of
the small variations of the specific evolutionary flux {\rm B},
variations of the observed $\alpha$ values indicate primarily different
values of $\tau_{PN}$ for PNs associated with the corresponding stellar
populations.  

Extended PN samples were collected for early-type galaxies (ETG) as
part of the Planetary Nebulae Spectrograph (PN.S, \cite[Douglas
  \etal\ 2002]{PNS02}) ETG
survey\\ (http://www.astro.rug.nl/$\sim$pns/). In ETGs, the PN number
density profile follows the surface brightness of the light
distribution, as shown for the PN.S galaxies (\cite[Coccato
  \etal\ 2009]{coccato09}) and for NGC~4697 (\cite[M\'endez
  \etal\ 2001]{M2001}), M~60 (\cite[Teodorescu \etal\ 2011]{teo2011})
and NGC~1399 (\cite[McNeil \etal\ 2010]{Mcneil+11}). Work is currently
ongoing for a sample of S0s galaxies (\cite[Cortesi
  \etal\ 2011]{Cort+11}; A. Cortesi, 2012 Ph.D thesis).

\begin{figure}
\begin{center}
 \includegraphics[width=3.1in]{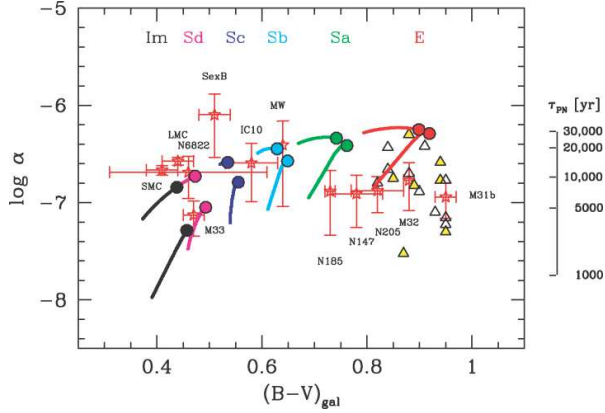} 
 \caption{A comprehensive overview of the luminosity-specific PN
   number in Local Group galaxies (open star symbols) and external
   ellipticals (solid and open triangle symbols). Superposed on the
   plot, there are template galaxy models. Galaxy evolution along
   the whole E-Sa-Sb-Sc-Sd-Im Hubble morphological sequence is tracked
   by models from 5 to 15 Gyr, with the latter limit marked by the big
   solid dots. Two model sequences are reported on the plot assuming
   an IFMR as from the standard case of a la Reimers mass-loss
   parameter $\eta$= 0.3 (lower sequence), and from the empirical
   relation of \cite[Weidemann (2000)]{W00} (upper sequence).  An
   indicative estimate of the mean representative PN lifetime (in
   years) is sketched on the right scale, according to
   Equation~\ref{taupn}. From \cite[Buzzoni
     \etal\ (2006)]{BAC2006}. \label{alphap}}
\end{center}
\end{figure}

{\underline{\it Planetary Nebulae trace stellar
    populations}}. \cite[Buzzoni \etal\ (2006)]{BAC2006} investigated
the time evolution of the luminosity specific PN number - the $\alpha$
parameter - using template galaxy models along the entire Hubble
morphological sequence E-Sa-Sb-Sc-Sd-Im. The results indicate that the
$\alpha$ parameter values predicted from these models depend on the
mass loss relation during the evolution from the turn-off of the
zero-age-main-sequence to the post-asymptotic giant branch phase; see
Figure~7 in \cite[Buzzoni \etal\ (2006)]{BAC2006}. When the $\alpha$
values are computed for the empirically calibrated Initial Final Mass
Relation (IFMR) of \cite[Weidemann (2000)]{W00}, there results a flatter
distribution from Im to E, see Figure~\ref{alphap}.

Figure~\ref{alphap} compares the $\alpha$ model values with
measurements from PN samples in ETGs and in galaxies from the Local
Group. While model predictions and observed values are consistent
for stellar populations in galaxy morphological types from Im
to Sa, the observed $\alpha$ parameters in ETGs with red and old
stellar populations are smaller than the predicted values by up to a
factor $7$, see also \cite[Hui \etal\ (1993)]{Hui+93} and
\cite[Ciardullo \etal\ (2005)]{ciardullo+05}.

\begin{figure}
\begin{center}
 \includegraphics[width=3.1in]{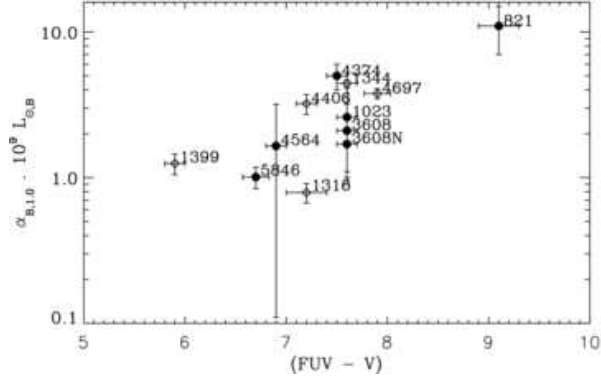} 
 \caption{Correlation between the $\alpha_{B,1.0}$ parameter and
   FUV$-$V color, measured from total extinction-corrected magnitudes
   FUV (from GALEX) and V (from RC3), from \cite[Coccato
     \etal\ (2009)]{coccato09}. Filled circles: sample galaxies for
   which $\alpha$ is calculated from the PN.S ETG sample data. Open
   diamonds: $\alpha$ values taken from \cite[Buzzoni
     \etal\ (2006)]{BAC2006}. \label{alphaFUV}}
\end{center}
\end{figure}

The relative PN deficiency in ETG supports the presence of a fraction
of low-mass cores with $M_{core} \leq 0.55 M_\odot$: for such low mass
cores, $\tau_{PN}$ may become shorter because the time required for
the excitation of the nebular envelope increases, and by the time it
happens, the nebular shell may be near to being lost. Furthermore the
part of the stellar population with $M_{core}\sim 0.52 M_\odot$ may
omit the PN phase entirely. These evolved stars may provide an
enhanced contribution to the hotter horizontal branch (HB) and post -
HB evolution as directly observed in M~32 and in the bulge of M~31,
which in turn implies that the far UV flux in ETGs should
anti-correlate with the measured $\alpha$ values. Such an
anti-correlation is indeed observed in the PN.S ETG sample, see
Figure~\ref{alphaFUV}.

A possible external mechanism that can act on a PN to shorten its
$\tau_{PN}$ is the interaction with the hot intracluster medium. The
ram pressure stripping by the hot high density gas in brightest
cluster galaxies (BCGs) coupled with the large velocity of a PN
through the medium can reduce the material in the PN envelope,
therefore effectively reducing $\tau_{PN}$ (\cite[Dopita
  \etal\ 2000]{dopmar00}, \cite[Villaver \& Stanghellini
  2005]{Vilsta05}).

{\underline{\it Planetary Nebulae populations trace distances}}. The
luminosity function of the [OIII] magnitudes associated with a PN
population can be described by the simple analytical formula given in
Eq.~\ref{PNLF}, whose absolute magnitude of the bright cut-off $M^*$
is observed to be invariant in different morphological types, from
late to early types (\cite[Ciardullo
  \etal\ 2002]{ciard+2002}). Because of the presence of a bright
cut-off, the PNLF can be effectively used as a secondary distance
indicator in both ETGs and late-type galaxies. One can ask whether the
invariance of the PNLF bright cut-off for populations of different ages
and metallicity can be explained on the basis of the evolution of
simple stellar populations.

From single star evolution, we know that a PN's peak flux in the
[OIII] $\lambda$ 5007\AA\ line is proportional to its core mass
(\cite[Vassiliadis \& Wood 1994]{vaswood94}), the core mass is
proportional to the turn-off mass (\cite[Kalirai
  \etal\ 2008]{kal+08}), and the turn-off mass decreases with
increasing age of the stellar population (\cite[Marigo
  \etal\ 2004]{marigo04}). Therefore the absolute magnitude of the
bright cut-off $M^*$ should become fainter in older and evolved
stellar populations. The observed invariance of the PNLF bright
cut-off indicates that the brightest PN may not be the end result of
simple evolution of single stars, but rather have different
progenitors, including blue stragglers (\cite[Ciardullo
  \etal\ 2005]{ciardullo+05}) and symbiotic stars (\cite[Soker
  2006]{soker06}).

\section{Planetary Nebulae as kinematical tracers}
In addition to measuring the $m_{5007}$ magnitude and determining the
distance to a given galaxy via the PNLF, the [OIII] $\lambda$ 5007
\AA\ emission allows us to measure of the line-of-sight (LOS) velocity
of a PN. By measuring $v_{LOS}$ for the PN population in the outer
regions of galaxies we can obtain the kinematics of the stellar
population at large radii and derive the projected two-dimensional
velocity field.

Since the work of \cite[Arnaboldi \etal\ (1994)]{arna+93} on NGC~1399
and of \cite[Hui \etal\ (1995)]{Hui+98} on Centaurus~A (NGC 5128), there
has been a considerable success in mapping the kinematics of the
stellar population in the outer regions of ETGs out to 5$R_e$
(\cite[Arnaboldi \etal\ 1996]{arna+96}, \cite[Arnaboldi
  \etal\ 1998]{arna+98}, \cite[M\'endez \etal\ 2001]{M2001},
\cite[Romanowsky \etal\ 2003]{rom+03}, \cite[Coccato
  \etal\ 2009]{coccato09}, \cite[Napolitano \etal\ 2009]{Nap+09},
\cite[McNeil \etal\ 2010]{Mcneil+11}, \cite[Napolitano
  \etal\ 2011]{Nap+11}, \cite[Teodorescu \etal\ 2011]{teo2011},
\cite[Cortesi \etal\ 2011]{Cort+11}), and intracluster light (ICL) in
nearby clusters (\cite[Arnaboldi \etal\ 2004]{arna+04}, \cite[Gerhard
  \etal\ 2005]{ger+05}, \cite[Gerhard \etal\ 2007]{ger+07},
\cite[Doherty \etal\ 2009]{Doh+09}, \cite[Ventimiglia
  \etal\ 2011]{ven+11}).

In the galaxies studied thus far, the PN density and kinematics is
everywhere consistent with the integrated light measurements within
the errors (\cite[Coccato \etal\ 2009]{coccato09}), with the possible
exception of NGC~4697 (\cite[Sambhus \etal\ 2006]{Sam+06}). The outer
halos kinematics from the PN samples in ETGs show the following
properties:
\begin{itemize}
\item there is a dichotomy in the outer halo kinematics, the velocity
  dispersion profile is either slowly falling or rapidly falling
  (\cite[Coccato \etal\ 2009]{coccato09}) with increasing radius.
\item The rotational properties of the halos traced by the ratio
  $v/\sigma$ where $v$ is the circular velocity and $\sigma$ is the
  velocity dispersion correlates with that within $R_e$ for the
  greater part of the sample galaxies observed so far.
\item The slow/fast rotators division in the outer halos
  (\cite[Coccato \etal\ 2009]{coccato09}) is similar as in the cores
  (\cite[Emsellem \etal\ 2007]{emsel+07}). There are some more
  complicated cases, as kinematic misalignment is more frequent at
  large radii (\cite[Shih \& M\'endez 2010]{ShMe10}, \cite[Teodorescu
    \etal\ 2011]{teo2011}).
\item A fraction of morphologically-undisturbed ellipticals show large
  distortions of their kinematics or presence of kinematic
  sub-components at large radii, as in the outer halos of NGC~1399
  (\cite[McNeil \etal\ 2010]{Mcneil+11}) and NGC~3311
  (\cite[Ventimiglia \etal\ 2011]{ven+11}).
\end{itemize}

\section{The brightest Planetary Nebulae in the Virgo Cluster }
The PN populations in the core of the Virgo cluster are studied via
narrow band imaging surveys, using the ``on-band, off-band''
technique, in which a field is imaged through a narrow band filter,
centered at the redshifted [OIII] $\lambda$ 5007 \AA\ emission in
these clusters, and a broad band filter. PN candidates are those
unresolved sources that have a color excess in [OIII] and no detected
continuum (\cite[Arnaboldi \etal\ 2002]{arna+02}). Tens of
intracluster PN (IPN) were discovered in the Virgo cluster, in its
core and in outer targeted fields (\cite[Arnaboldi
  \etal\ 2002]{arna+02}, \cite[Okamura \etal\ 2002]{Oka+02},
\cite[Arnaboldi \etal\ 2003]{arna+03}, \cite[Feldmeier
  \etal\ 2003]{Feld+03}, \cite[Feldmeier \etal\ 2004]{Feld+04},
\cite[Aguerri \etal\ 2005]{aguer+05}, \cite[Castro-Rodriguez
  \etal\ 2009]{castro+09}). Such surveys are complete 0.5 mag down the
PNLF, therefore they sample mostly the brightest PNs in these stellar
populations.  The spectroscopic follow-up with multi-object
(\cite[Arnaboldi \etal\ 2003]{arna+03}) and multi fiber spectrographs
(\cite[Arnaboldi \etal\ 2004]{arna+04}, \cite[Doherty
  \etal\ 2009]{Doh+09}) allow 
\begin{itemize}
\item to detect the second [OIII] $\lambda$
4959 \AA\ emission besides $\lambda$ 5007 \AA\ in single spectra,
\item measure the radial velocities and associate PNs
with the dynamical components along the LOS, 
\item determine PN physical parameters, the $m_{5007}$ and the
  velocity $v_{EXP}$ derived from the FWHM of the [OIII] $\lambda$
  5007\AA\ emission, i.e. the expansion velocity of a PN nebular
  envelope. These PN physical parameters can be correlated with those
  of its parent stellar population (\cite[Arnaboldi
    \etal\ 2008]{arna+08}).
\end{itemize}

\begin{figure}
\begin{center}
 \includegraphics[width=3.1in]{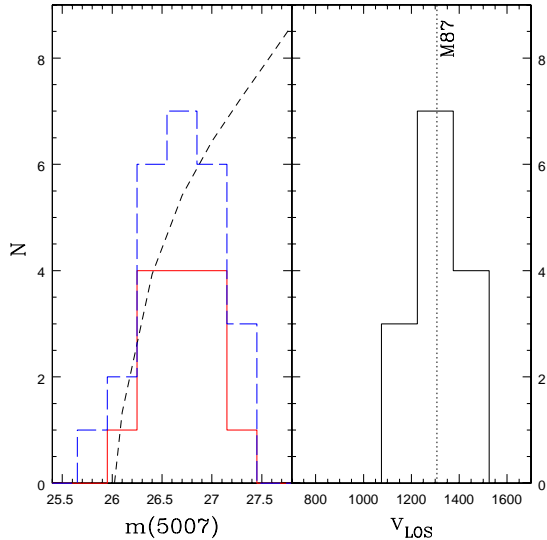} 
 \caption{Left panel: the observed PNLF in the Virgo cluster core
   fields, in 0.3 mag bins. Long dashed blue line: PNLF for the entire
   spectroscopically confirmed sample of PNs in the Virgo cluster
   core. Red continuous line: PNLF for the M87 halo PNs. Short dashed
   black line: analytical formula for the PNLF with $m^* =
   26.0$. Right panel: LOS velocity distribution for the 14 M87 halo
   PNs. The systemic velocity of M87 is also indicated. From
   \cite[Arnaboldi \etal\ (2008)]{arna+08}. \label{M87pnlf}}
\end{center}
\end{figure}

In case of M87, the BCG in the Virgo cluster, the PN LOS velocity
distribution traces the M87 halo out to 150~kpc and shows that the
halo PN population has a slightly brighter cut-off than that in the
galaxy's central region, by $0.3-0.4$ mag, see the observed M87 PNLF
in Figure~\ref{M87pnlf}. Because the LOS distance to these PNs is
known, via the physical association with the M87 halo, their absolute
magnitude is also known, and so are then the core masses required to
power the detected [OIII] $\lambda$ 5007 \AA\ fluxes: for the
brightest PN in the M87 halo, the core masses are $\simeq 0.62
M_\odot$. The IFMR for solar metallicity stars predicts that PN
progenitors with $2.2 M_\odot$ give final core masses of $0.62
M_\odot$. Turnoff masses of $\simeq 2 M_\odot$ belong to populations
with ages $\sim 1 Gyr$. It is unlikely that such progenitors exist in
the Virgo core, as the stellar population is mostly old (\cite[Williams
  \etal\ 2007]{wills+07}): therefore close binaries and blue
stragglers are indicated as most likely progenitors for these
intrinsically bright PNs.

\section{The brightest Planetary Nebulae in the Hydra I Cluster }

{\underline{\it The Multi-Slit Imaging Spectroscopy Technique}}.
The nearest clusters beyond Virgo and Fornax are at distances of about
50 Mpc, and the brightest PN in the Hydra~I and the Centaurus clusters
have fluxes of $8 \times 10^{-18}$ erg s$^{-1}$ cm$^{-2}$. They cannot be
detected using standard narrow band imaging because the noise in the
$40 - 60$ \AA\ sky is of the same order of the signal we want to
detect.

A successful technique to detect the [OIII] $\lambda$
5007\AA\ emission from the brightest PNs in these clusters is the
``Multi-Slit Imaging Spectroscopy technique'' (MSIS, \cite[Gerhard
  \etal\ 2005]{ger+05}, \cite[Arnaboldi \etal\ 2007]{arna+2007}). This
is a blind search technique that combines a mask of parallel multiple
slits with a narrow band filter, centered at the redshifted [OIII]
emission of the PNs in these clusters, plus a dispersing element to
obtain spectra of all PNs that lie behind the slits. The sky noise at
the emission line of a PN comes from a few \AA\ only, depending on
slit width and spectral resolution (\cite[Gerhard
  \etal\ 2005]{ger+05}).

\begin{figure}
\begin{center}
 \includegraphics[width=3.1in]{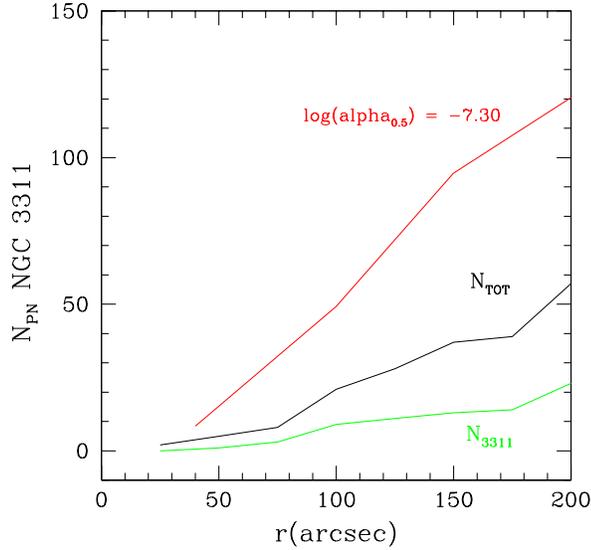} 
 \caption{Observed and predicted cumulative PN number for the NGC~3311
   halo and the total (NGC~3311 plus ICL) light in the Hydra~I core,
   as function of the radial distance from the center of NGC~3311. Red
   line: predicted cumulative number of PNs computed using $\log\alpha
   = -7.30$. Green line: measured cumulative number of PNs associated
   with the NGC~3311 halo on the basis of the $v_{LOS}$. Black line:
   measured cumulative number of PNs corresponding to the total light
   in the Hydra~I core. From \cite[Ventimiglia
     \etal\ (2011)]{ven+11}. \label{alpha3311}}
\end{center}
\end{figure}

{\underline{\it PN population in the halo of NGC~3311}}. The MSIS
observations of the Hydra~I cluster carried out with the FORS2
spectrograph at the VLT detected 56 PNs and 26 background galaxies in a
$6'.8 \times 6'.8$ field centered on the BCG NGC~3311 in the cluster
core (\cite[Ventimiglia \etal\ 2011]{ven+11}). In addition to the
study of the halo kinematics, \cite[Ventimiglia \etal\ (2011)]{ven+11}
measured the luminosity specific PN number $\alpha$ for the stellar
population in the extended NGC~3311 halo, and obtain a value that
is a factor $4-6$ lower than the predicted value of $\log\alpha =-7.30$
from the $\alpha$ vs. $FUV - V$ color correlation for the ETG
populations shown in Figure~\ref{alphaFUV}. In Figure~\ref{alpha3311}
the observed cumulative PN number for the NGC~3311 halo (green line)
and the total NGC~3311 plus ICL light (black line) in the Hydra~I core
are plotted together with the predicted cumulative PN number for the
NGC~3311 plus ICL (using $\log\alpha = -7.30$, red line), as function
of the radial distance from the center of NGC~3311. The large
discrepancy between the red and green lines indicates a decrement of
PNs in the NGC~3311 halo. This PN deficit may signal that additional
effects are at work that shape the properties of the PN population in
the NGC~3311 halo.

A possibility is that ram pressure against the hot X-ray emitting gas
in the halo of NGC~3311 is high enough to severely shorten $\tau_{PN}$,
the visibility time of a PN, see Section~\ref{intro}.  With a density
of the intracluster medium (ICM) inside $5'$ around NGC~3311 of about
$\rho_{ICM} \simeq 6 \times 10^{-3}$ cm$^{-3}$ and a typical velocity
of $v=\sqrt3 \times 450$ kms$^{-1}$ $\sim 800$ kms$^{-1}$, the ram
pressure on the NGC~3311 halo PNs is $\sim 40$ times stronger than in
the simulated case by \cite[Villaver \& Stanghellini (2005)]{Vilsta05}
for the Virgo cluster. The ram pressure effect could be much stronger
in the Hydra~I cluster and shorten significantly $\tau_{PN}$ for the
NGC~3311 halo population.

\section{Summary and Conclusions }

In this review we illustrate that PN populations are ubiquitous to the
light associated with external galaxies and the diffuse light in the
cores of nearby clusters. Deep photometry and kinematics of PN
populations show that they trace star light and motions in the extended,
luminous halos of BCGs and ICL.

The PN population shows dependencies on the stellar population age and
metallicity. Recent observations of PNs in the Hydra~I cluster
indicate that there may be an additional effect by the hot ICM, which
could decrease the visibility lifetime $\tau_{PN}$ of PNs. Further
theoretical and observational work is needed to quantify effects of
stellar population age, metallicity and ICM on both the PNLF bright
cut-off and the luminosity specific PN number $\alpha$.

The study of the PNs in the ICL and the halos of galaxies is important
for constraining how and when these components formed. In the case of
the Hydra~I cluster the results indicate that the build-up of the BCG
halo and the nearby ICL continues, and that the formation of the outer
halo and ICL is an on-going and long lasting process.

\section{Acknowledgments}
I wish to thank my collaborators for their enthusiasm and support and
the organizers for inviting me to give a review on PN populations in
external galaxies at the IAU Symposium 283.\\

\end{document}